\documentclass[prb,showpacs,floatfix,superscriptaddress,amsmath,amssymb,twocolumn]{revtex4-1}
\usepackage{graphicx}
\usepackage{graphics}
\usepackage{dcolumn}
\usepackage{bm}
\usepackage{pstricks}
\usepackage{amssymb}
\usepackage{hyperref}
\usepackage{nccmath}

\definecolor{lineadecolor}{rgb}{0.35,0.5,0.6}
\definecolor{ddcol}{rgb}{0.8,0.1,0.1}
\definecolor{subsectioncolor}{rgb}{0.1,0.01,0.5}

\definecolor{celeste}{rgb}{0.8,0.87,0.99}

\def\nn{\nonumber}
\newcommand{\ba}{\begin{eqnarray}}
\newcommand{\ea}{\end{eqnarray}}
\def\be{\begin{equation}}
\def\ee{\end{equation}}

\def\spin{\vec{\mathbf{S}}}

\begin{document}

\title{Magnetization process in a frustrated plaquette dimerized ladder.}

\author{ F.\ Elias}
\affiliation{IFLP - CONICET. Departamento de F\'isica, Facultad de Ciencias Exactas. Universidad Nacional de La Plata,C.C.\ 67, 1900 La Plata, Argentina.}

\author{ M.\ Arlego}
\affiliation{IFLP - CONICET. Departamento de F\'isica, Facultad de Ciencias Exactas. Universidad Nacional de La Plata,C.C.\ 67, 1900 La Plata, Argentina.}

\author{ C.A.\ Lamas}
\affiliation{IFLP - CONICET. Departamento de F\'isica, Facultad de Ciencias Exactas. Universidad Nacional de La Plata,C.C.\ 67, 1900 La Plata, Argentina.}

\begin{abstract}
The magnetic phase diagram of a plaquette dimerized antiferromagnetic system is studied by using a combination of numerical and analytical techniques.
For the strongly frustrated regime, series expansions and bond operators techniques are employed to analyze zero magnetization plateau, whereas low energy effective models are used to study the complete magnetization process.  \\
The interplay between frustration and dimerization gives rise to a rich plateaus structure that is captured by effective models and corroborated by numerical density matrix renormalization group simulations, in particular the emergence of intermediate plateaus at M = 1/4 and 3/4 of saturation in the magnetization curve.

\end{abstract}

\maketitle

\section{Introduction}

Antiferromagnetic spin ladders are paradigmatic examples of quantum magnetism in low dimensions \cite{Dagotto-Rice-Science,Mila-2016,mila-2016-bis,mila-berry-phase,laflorencie-2011,Totsuka1,NH4CuCl3-magnetization,NH4CuCl3-model,cabra-2001,Mila-magnons-spinons}.
These systems exhibit a plenty of interesting properties, such as absence of long range order, energy gaps, magnetic disordered phases, among others \cite{Mikeska-ch1}.
One particular issue of these systems that captured both experimental and theoretical efforts is the existence of magnetization plateaus \cite{Mila-capitulo-plateaus}. For a quantum spin-S chain, the necessary condition for the occurrence of a magnetization plateau has been established by Oshikawa, Yamanaka and Affleck (OYA)\cite{OYA} as
\ba
N(S-m) \in \mathbb{Z},
\ea
where  $N$ is the period of the spin state, $S$ the magnitude of spin and $m$ the magnetization per site in units of $g \mu_B$.
The OYA criterium has been successfully applied to different one dimensional magnets \cite{Lamas-Pujol_2010,cabra-2001} and revisited by different techniques\cite{TTH,totsuka-2015,topological-condition,plat-2015}.
In this context several  quasi-one-dimensional compounds has been studied in the past years. As an example we mention the compound  Cu$_2$Cl$_4\cdot$D$_8$C$_4$SO$_2$, which presents a singlet ground state. In order to describe the properties of this material different models have been proposed \cite{Fujisawa-2003,experimental_tube_prb,tube_inelastic_scatering}, however the magnetic properties of this compound has not been completely understood.
Another system with a remarkable magnetic behavior is the $S = \frac12$ zig-zag ladder compound NH$_4$CuCl$_3$. In this material the magnetization curve presents two plateaus at $m =\frac14 $ and $m =\frac34 $ of the saturation magnetization, irrespective of the external field direction, but not at $m =0 $ and $m =\frac12 $ \cite{NH4CuCl3-magnetization}. 

In this paper we study an antiferromagnet model in the zig-zag geometry showing that the interplay between geometrical frustration and plaquette dimerization promotes the existence of plateaus at $m =\frac14 $ and $m =\frac34 $.
Although we do not intend to provide a theoretical description of these materials, our results may be relevant for the discussion of the emergence of rational magnetization plateaus and its magnon and spinon excitations.
The model analyzed in this work belongs to a family of frustrated antiferromagnetic systems with a fully dimerized
exact ground state \cite{Brenig-tetrahedral,Matera-Lamas_2014,Lamas-Matera_2015}. Recently it was shown that by imposing certain \emph{exact} condition to the couplings, the exact ground state of the system is robust against the inclusion of weak disorder \cite{Lamas-Matera_2015}. This property provides to the model interesting magnetic properties when the unit cell is increased in order to consider plaquette dimerization.
The exact condition can be imposed both in one or the two different square plaquettes giving a rich plateaus structures, an aspect that will be exploited in this work.

The outline of the paper is as follows. In Section \ref{sec:exgs} we study the zero magnetic field phase diagram. Starting form the line where the ground state is exactly determined we use a bond operator formalism to study the excitation gap. In Section \ref{sec:PlateauPhases} we analyze the effect of an external magnetic field on the model. We study the magnetization process by means of density matrix renormalization group (DMRG) calculations and exploring the parameters space we determine regions where different plateaus are present.
In the highly dimerized plaquette regime we derive low energy effective models and studying the excitations we estimate the critical magnetic fields and the plateau weights. The interplay between dimerization and frustration results in a rich plateaus structure that is captured by the low energy model and is consistent with the numerical results obtained through DMRG.
Finally in Section \ref{sec:conclusions} We present the conclusions. Some relevant formulas and expressions used throughout the work are shown in the Appendix.

\section{Zero magnetic field phase diagram}
\label{sec:exgs}
\subsection{Exact ground state in the absence of magnetic field}
\label{sec:exact-lines}

We consider the following Heisenberg model on a two legs spin-$S$ ladder
\begin{figure}[t]
\includegraphics[width=0.8\columnwidth]{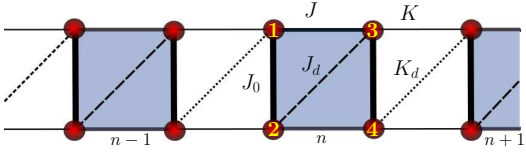}
\caption{(Color online). Schematic representation of the dimerized zig-zag ladder. }
\label{fig:ladder}
\end{figure}
\ba
\label{eq:general_Hamiltonian}
\mathbf{H}&=&\sum_{n} J_{0} \left(\spin_{1,n}\cdot\spin_{2,n}+\spin_{3,n}\cdot\spin_{4,n}\right)\\\nn
&+&J \left(\spin_{1,n}\cdot\spin_{3,n}+ \spin_{2,n}\cdot\spin_{4,n}\right)+J_{d} \, \spin_{2,n}\cdot\spin_{3,n} \\\nn
&+& K\left( \spin_{3,n}\cdot\spin_{1,n+1}+\spin_{4,n}\cdot\spin_{2,n+1}\right)+K_{d}\, \spin_{4,n}\cdot\spin_{1,n+1},
\ea
where $\spin_{j,n}$ represents the $j$-th spin on the unit-cell $n$. This ladder is schematized in Fig. \ref{fig:ladder}.
Following the lines scketched in Ref. \onlinecite{Matera-Lamas_2014} we can show that, provided the conditions $J_{d}=2J$ and $K_{d}=2K$,
the ground state of the system corresponds to the fully dimerized state
$ |{\psi}\rangle=\bigotimes_{i=1}^{N_{r}}|s\rangle_{i}$, with
\begin{eqnarray}
\label{eq:dimerstate}
 |s\rangle_{i}= \frac{1}{\sqrt{2 S+1}}\sum_{m=-S}^{S}(-1)^{m+S}|m,-m\rangle_{i},
\end{eqnarray}
where the index $i$ label the rungs in the ladder and $N_{r}$ is the number of rungs. The state
$|m,-m\rangle_{i}$ is a product state such that ${\bf S}^{z}_{2 i-1,n}|m,-m\rangle_{i}=-{\bf S}^{z}_{2 i,n}|m,-m\rangle_{i}=m |m,-m\rangle_{i}$
where the index $i=1,2$ corresponds to the two rungs in the unit-cell $n$.
This state remains the ground state of the system in the range of couplings $J<J_{0}/2$ and $K<J_{0}/2$ .
For this purpose, it is convenient to rewrite the Hamiltonian in the case $J_d=2J$, $K_d=2K$ as follows
\begin{eqnarray}
\mathbf {H}&=&\sum_{n}(\mathbf{h}_{J}(n)+\mathbf{h}_{K}(n)),
\end{eqnarray}
where $\mathbf{h}_{J}(n)$ and $\mathbf{h}_{K}(n)$  are Hamiltonians of the square plaquettes on unit-cell $n$,
given by
%
\ba
\label{eq:hJ}
\mathbf{h}_{J}(n)&=&\frac{J}{2}\left[ \left(  \spin_{1,n}+\spin_{2,n}+\spin_{3,n}  \right)^{2} \right. \\\nn
&+& \left. \left(  \spin_{2,n}+\spin_{3,n}+\spin_{4,n}  \right)^{2}    \right]\\ \nn
&+&\frac{(J_{0}-2J)}{4}\left[ \left( \spin_{3,n}+\spin_{4,n}  \right)^{2}   + \left(  \spin_{1,n}+\spin_{2,n} \right)^{2}    \right]\\ \nn
&-&\frac34(J_{0}+J),
\ea
\ba
\label{eq:hK}
\mathbf{h}_{K}(n)&=&\frac{K}{2}\left[ \left(  \spin_{3,n}+\spin_{4,n}+\spin_{1,n+1}  \right)^{2} \right. \\\nn
&+& \left.  \left(  \spin_{4,n}+\spin_{1,n+1}+\spin_{2,n+1}  \right)^{2}    \right]\\ \nn
&+&\frac{(J_{0}-2K)}{4}\left[ \left( \spin_{3,n}+\spin_{4,n}  \right)^{2}   + \left(  \spin_{1,n}+\spin_{2,n} \right)^{2}    \right]\\ \nn
&-&\frac34(J_{0}+K).
\ea
In this way, is easy to see that $\mathbf {H}$ is a semi-definite positive operator if $J<J_{0}/2$ and $K<J_{0}/2$. In the case $S=1/2$, the energy per square plaquette is bounded from bellow by
\small
\ba
\frac{E_{gs}}{N_s}&\geq&\min_{l_{1,2},l_{3,4}=0,1}   \left\{ -\frac34(J_{0}+J)-\frac34(J_{0}+K)  \right. \\\nn
&+&\frac{J}{2}\sum_{n=1}^2\left[ |\frac12-l_{2n-1,2n}|(|\frac12-l_{2n-1,2n}|+1)\right]\\\nn
&+& \left. \frac{(J_{0}-2J)}{4}\sum_{n=1}^2\left[ l_{2n-1,2n}(l_{2n-1,2n}+1) \right] \right\}\\\nn
&+&\frac{K}{2}\sum_{n=1}^2\left[ |\frac12-l_{2n-1,2n}|(|\frac12-l_{2n-1,2n}|+1)\right]\\\nn
&+& \left. \frac{(J_{0}-2K)}{4}\sum_{n=1}^2\left[ l_{2n-1,2n}(l_{2n-1,2n}+1) \right] \right\},
\ea
\normalsize
%
where $l_{1,2}$ ($l_{3,4}$) is the total spin in the rung between sites $1$ and $2$ ($3$ and $4$). It is straightforward to see that the minimum value correspond to $l_{1,2}=l_{3,4}=0$, given the bounding condition
\ba
\frac{E_{gs}}{N_s}\geq -\frac32 J_{0}.
\ea
To conclude, writing the Hamiltonian in terms of local operators in each rung
\ba
\vec{\mathbf{L}}_{1,2}(n)&=&\spin_{1,n}+\spin_{2,n}\\\nn
\vec{\mathbf{L}}_{3,4}(n)&=&\spin_{3,n}+\spin_{4,n}\\\nn
\vec{\mathbf{K}}_{1,2}(n)&=&\spin_{1,n}-\spin_{2,n}\\\nn
\vec{\mathbf{K}}_{3,4}(n)&=&\spin_{3,n}-\spin_{4,n},
\ea
%
is easy to see that
if we restrict the couplings to satisfy the condition $J_d=2J$ and $K_d=2K$, the state $|{\psi}\rangle$ is an eigenstate with eigenvalue $E=-\frac32 J_{0} N_s$.
Therefore we have proved that the state $|{\psi}\rangle$ is the ground state of the system along the lines  $J_d=2J$ and $K_d=2K$ as long as the couplings satisfy  $J<J_{0}/2$ and $K<J_{0}/2$. Notice that these two last constraints represent a sufficient condition for the existence of this singlet ground state, but the true range of couplings can be larger.

This exact result is a very convenient starting point for dimer expansions. Notice that the existence of a line where the singlet product state is the true ground state of the system allows us to use dimer expansions around a point in the parameter space where the Hamiltonian is maximally frustrated. In the following Section we present some results obtained by different techniques based on dimer expansions.

\subsection{Bond Operators appproach}

Close to the line in the parameter space where the ground state corresponding to Hamiltonian \ref{eq:general_Hamiltonian} is exactly known it is convenient to exploit  the description in terms of bosonic operators, the so called \emph{bond operators} (BO) \cite{SachdevBO}
which label the dimer's singlet-triplet spectrum.
On the exact line the ground state is a direct product of states on each square plaquette. On each plaquette the state corresponds to a product of a singlet between spins located at sites  $1,n$ and $2,n$ and another singlet
between spins at $3,n$ and $4,n$. Within BO theory the four spins $\vec{S}_{i}$ on each
square plaquette are expressed as

\begin{fleqn}
\begin{align}
\label{eq:BO-A}
S_{\stackrel{{\scriptstyle 1,n}}{{\scriptstyle 2,n}}}^{\alpha}&=\frac{1}{2}(\pm
s_{A,n}^{\dagger}a_{\alpha,n}\pm a_{\alpha,n}^{\dagger}s_{A,n}-\sum_{\beta,\gamma}i
\varepsilon_{\alpha\beta\gamma}a_{\beta,n}^{\dagger}a_{\gamma,n}^{\phantom{\dagger}})\\
\label{eq:BO-B}
S_{\stackrel{{\scriptstyle 3,n}}{{\scriptstyle 4,n}}}^{\alpha}&=\frac{1}{2}(\pm
s_{B,n}^{\dagger}b_{\alpha,n}\pm b_{\alpha,n}^{\dagger}s_{B,n}-\sum_{\beta,\gamma}i
\varepsilon_{\alpha\beta\gamma}b_{\beta,n}^{\dagger}b_{\gamma,n}^{\phantom{\dagger}}),
\end{align}
\end{fleqn}
where $s_{A,n}^{(\dagger)}$and $a_{\alpha,n}^{(\dagger)}$ destroy(create)
the singlet and triplet states of the dimer between sites $1,n$ and $2,n$ and greek labels, $\alpha=1,2,3$,
refer to the threefold triplet multiplet. Equivalently operators $s_{B,n}^{(\dagger)}$and $b_{\alpha,n}^{(\dagger)}$ act on the dimer between sites $3,n$ and $4,n$. A hard-core constraint
\ba
\label{eq:constraint-A}
\displaystyle{s_{A,n}^{\dagger}s_{A,n}}+\sum_{\alpha}a_{\alpha,n}^{\dagger}
a_{\alpha,n}^{\phantom{\dagger}}&=&1,\\
\label{eq:constraint-B}
s_{B,n}^{\dagger}s_{B,n}+\sum_{\alpha}b_{\alpha,n}^{\dagger}
b_{\alpha,n}^{\phantom{\dagger}}&=&1,
\ea
is implied on each sublattice, which renders the algebra of the r.h.s of Equations. (\ref{eq:BO-A}) and (\ref{eq:BO-B})identical to that of spins.
Inserting the BO representation into a spin model leads to an interacting
Bose gas. In the BO-MFT,  singlets are condensed by
$s^{(\dagger)}\rightarrow s\in\,$Re and the constraint in the number of bosons per site is
satisfied on average with a global Lagrange multiplier $\lambda$. In this approach
terms only up to second order in the BOs are retained and the quadratic Hamiltonian can be diagonalized by standard Bogoliubov transformation leading to an energy $E$ \emph{per unit cell} of
\begin{equation}
E={-}\frac{3}{4}J_{0}-\frac{3}{2}s^{2}J_{0}+(5-2s^{2})\lambda+\frac{3}{N}\sum_{k}(\omega^{+}_{k}+\omega^{-}_{k}),\label{wa8}
\end{equation}
with the \emph{triplon dispersion}
\begin{equation}
\label{eq:BO-dispersion}
\omega^{\pm}_{k}=|a|\sqrt{1\pm|g|\frac{s^{2}}{a}\beta_{\delta}(k)},
\end{equation}
where
\begin{eqnarray}
\beta_{\delta}(k)&=&\frac{1}{2}\sqrt{1+\delta^2 + 2\delta \cos(k)},
\end{eqnarray}
$a=\frac{J_{0}}{4}-\lambda$, $g=J_d-2J$, and $s=\langle s_{A,n} \rangle=\langle s^{\dagger}_{A,n} \rangle=\langle s_{B,n} \rangle=\langle s^{\dagger}_{B,n} \rangle$ is a real parameter and we have restricted the study
to the case $K=\delta J$ and $K_d=\delta_d J_d$ with $\delta = \delta_d$ \emph{ie} \emph{homogeneous} dimerization. The general case will be analyzed with other techniques in the following Sections.
In order to obtain the mean field parameters $s$ and $a$  the energy $E$ is extremized, obtaining two
self-consistency equations $\partial E/\partial a=0$ and $\partial E / \partial s=0$.
These two equations can be combined obtaining
\small
\ba
\label{eq:BO-selfcon1}
d&=&\frac{5}{2J_{0}}-\frac{3}{8 \pi J_{0}}\hbox{sign}(a)\int dk \left( \frac{1}{\gamma_-(k)} +\frac{1}{\gamma_+(k)} \right),\\
\label{eq:BO-selfcon2}
a&=&J_{0}-\frac{3}{16\pi}|g|\hbox{sign}(a) \int dk  \left( \frac{\beta_{\delta}(k)}{\gamma_+(k)} -\frac{\beta_{\delta}(k)}{\gamma_-(k)} \right),
\ea
\normalsize
where $\gamma_\pm(k)=\sqrt{1\pm d |g| \beta_{\delta}(k)}$ and $d=s^{2}/a$. These equations can be solved sel-consistently.

We mention in passing, that the trivial limit, i.e. $g=0$, leads to $d=1/J_{0}$, $s=1$, and $\lambda=-\frac{3}{4}J_{0}$, and therefore to a \emph{singlet-triplet gap} of $\Delta=1$ and a ground state energy of
$E/N={-}\frac32 J_{0}$, which is consistent with two saturated singlets per unit cell.

\subsection{Triplet dispersion and energy gap}

In this Section we will analyze the triplet dispersion in absence of an external magnetic field in the model given by Eq.(\ref{eq:general_Hamiltonian}). For this, we will compare the results obtained by applying techniques of different nature, which complement each other and bring a perspective that none of the techniques can provide separately. The methods used here are on the one hand the numerical technique DMRG, and on the other hand strong coupling dimer and plaquette expansions, as the BO method from the previous Section, and the series expansions technique (SE) considered in the Appendix. \\
The Fig. \ref {fig:gap-BO} summarizes the main result of this Section, obtained by the different techniques mentioned previously, for the triplet gap, \emph{ie} the minimum of the dispersion, as a function of the parameter $ g = J_d-2J $. The case considered in this figure is that of a homogeneous structure without dimerization, \emph{ie} $\delta = \delta_d=1$. The DMRG results, shown with blue dots in Fig. \ref {fig:gap-BO}, for $ J = 0, 0.1, 0.2, 0.3 $ and 0.4 from right to left, and were calculated on finite structures of $ L = 40 $ plaquettes (number of sites $N_s=160$) with periodic boundary conditions (PBC), maintaining up to $ M = 520 $ states in the computation, which has shown to be enough to achieve the required precision. Unless specified otherwise, in the remainder of the work these will be the DMRG parameters used. All the DMRG computations presented in  this work were performed with the open-source code ALPS \cite{ALPS}. \\
On the other hand, the SE results are indicated by red line in this figure with the same $J-$parametrization as DMRG, and were obtained from an expansion in dimers to $O(10)$ by using the method of continuous unitary transformations (CUT) \cite{SE-CUT}. Finally, the BO results are of two types. The first one is the \emph{Holstein-Primakoff} approximation (BO-HP), which is indicated by green line in the Fig. \ref {fig:gap-BO}. In the BO-HP the self-consistent parameters in Eqs.(\ref{eq:BO-selfcon1},\ref{eq:BO-selfcon2}) are solved for $g=0$ (obtaining $s=a=d=1$) and are replaced in the dispersion Eq.(\ref{eq:BO-dispersion}) which, after the re-scaling $k \rightarrow 2k$, imposed by halving the unit cell (dimer basis), takes the simple form $\omega_{HP}=\sqrt{1-|g|\cos k}$. Note that this expression coincides with the treatment of the model under the random phase approximation (RPA) method \cite{link-BO-SE}. The second type of BO result is the self-consistent solution of mean-field BO, Eqs.(\ref{eq:BO-selfcon1}
,\ref{eq:BO-selfcon2}), and is indicated by the orange line in the Fig. \ref {fig:gap-BO}. \\

\begin{figure}[t]
\includegraphics[width=0.9\columnwidth]{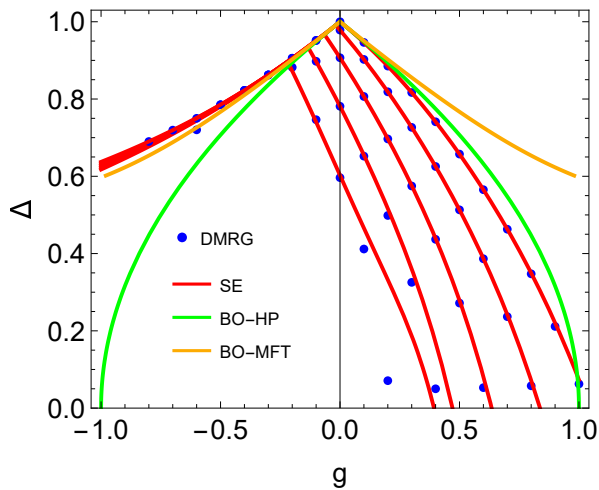}
\caption{(Color online) Spin gap $\Delta$ versus $g=J_d-2J$ obtained by means of different techniques employed in the work. The SE (red lines) and DMRG (blue dots) results, are parameterized for $ J = 0, 0.1, 0.2, 0.3 $ and 0.4 from right to left. Notice that for bond operators approximations, Holstein-Primakoff (BO-HP, green line) and self-conistent (BO-MFT, orange line), the spin gap (and the dispersion) depend only on $g$, which is an artifact of the approximation. The peak shape of the gap maxima indicates a crossing of levels between  $k = 0$ (right) and  $k = \pi$ (left) dispersion branches, detected by SE. Gap right branches together with the peaks shift to the left by increasing $J-$values and tend to collapse into a single g-dependent line for $g <0$, where the methods (except BO-HP) exhibit a very good agreement.}
\label{fig:gap-BO}
\end{figure}
Several comments are in order. First of all note that, apart from the $g-$dependence, exists a \emph{residual} interaction, represented here through the $J-$parametrization, that is captured by SE and DMRG, but not by BO.
That is, BO encapsulates the entire dependence on $J$ and $J_d$ in $g$, which is artifact of the mean field approximation. \\
In addition, as can be observed, there is a very good quantitative agreement between DMRG and SE results, which is maintained until intermediate values of $ J $, according to the perturbative character of the SE approximation.\\
On the other hand, note that the absolute maximum of the gap $(= 1)$ at $g = 0$, as well as the other maxima in Fig. \ref {fig:gap-BO}, are non-derivable (peaks), representing a crossing of the dispersion branches at $k = 0$ (right) and $k=\pi$ (left) of the peak, which are detected by the SE. This in turn reflects a change from a dimerized chain-type to ladder-type of spectrum, respectively.\\
Regarding the role of the residual interaction, it shifts the gap peak to the left $(g<0)$. However its effect on the two dispersion branches is very different. While the branch $k = 0$ (to the right of the peak) moves along with the peak to the left by increasing $J$, the branch $k = \pi$ on the left tends to collapse on a single curve, \emph{ie} its dependence on $J$ and $J_d$ is captured mainly by $g$. These features are detected by means of SE and numerically corroborated by DMRG as can be seen in Fig. \ref {fig:gap-BO}.\\
Finally, with respect to BO, it is worth noting that although the mean field level does not take into account the residual interaction, but only g, it is able to capture some essential aspects of the gap behaviour, not only qualitatively, but also up to a certain quantitative level. Here the two approaches BO describe the gap with different precision on both sides of the gap peak. As can be seen in Fig. \ref {fig:gap-BO}, the BO-HP (green line) adequately describes the right branch of the gap (in particular the $J=0$ line), while the BO-MFT (orange line) agreement for the left branch is very good, as compared with SE and DMRG results. This tendency of the dispersions to collapse with a single $g-$dependence for $g <0$ could be related to the fact that in that region the gap remains large, ie more adiabatically connected to the exact line $g = 0$, as shown in the next Section through a sweep in the $J-J_d$ plane of the gap. Finally let us mention that in quantitative terms, the role of the residual 
interaction and the dependence on $g$ is clarified in the Appendix, where it is explicitly shown that at leading order the dispersions obtained by SE and BO-HP coincide and depend only on $g$.

\section{Magnetic field phases}
\label{sec:PlateauPhases}
\subsection{Plateau phases without plaquette dimerization }
\begin{figure}[t]
\includegraphics[width=0.8\columnwidth]{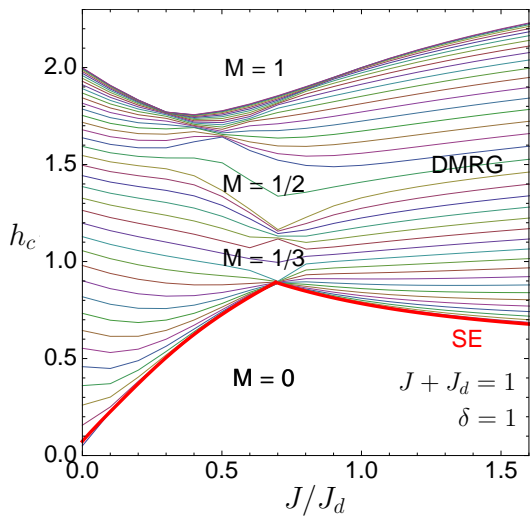}
\caption{Critical fields (thin-lines) delimiting the different magnetization sectors on a finite system, determined by DMRG, as a function of $ J / J_d $, along the line $ J + J_d = 1 $. Note the presence of two intermediate plateaus at $M=1/3$ and 1/2. The bold red line indicates $M=0$ critical line (in thermodynamic limit), where SE triplet gap vanishes, showing an excellent agreement with the corresponding DMRG determination.}
\label{fig:phases-hc-delta-1}
\end{figure}

In this Section we will analyze the effect of an external magnetic field on the model given by Eq.(\ref{eq:general_Hamiltonian}). To simplify the analysis we will start by studying the case of homogeneous frustrated plaquettes, \emph{ie} $ \delta = \delta_d = 1 $. Although this case has been previously studied in different regions of the space of parameters and employing different techniques \cite{Shastry-Sutherland81,Tonegawa87,Chitra95,White96,Mikeska96,Yokoyama99}, our analysis presents it in a unified way and contextualizes our subsequent study of the general case with frustration and dimerization between-plaquettes of the following Sections.\\
As it is known the presence of a magnetic field opens the possibility of the emergence of plateau phases in the magnetization curves, where the magnetization is fixed at a certain value, for a finite interval of the magnetic field.\\
In Fig. \ref{fig:phases-hc-delta-1} we show the critical fields $h_c$ delimiting the different possible values of magnetization, determined by DMRG, in function of $ J / J_d $, along the line $ J + J_d = 1 $. The choice of the parameters in this figure is in order to show in a single representation the emergence of intermediate plateaus. As can be observed, apart from the zero field $(M=0)$ and saturation $(M=1)$ plateaus, two additional plateaus at $M=1/3$ and 1/2 are present. Note that the critical line crossing each of this two plateaus represents a small jump that is an artifact due to the open boundary conditions used for the computation.\\
Additionally, in Fig. \ref{fig:phases-hc-delta-1}, the critical line delineating the plateau at $M = 0$, obtained by the SE triplet gap closure condition, is presented with red line. The peak of this curve represents the crossing of the gap closing branches at $ k = 0 $ and $ k = \pi $. As can be observed, the quantitative agreement between DMRG and SE for the line $ M = 0 $ is excellent throughout the range of parameters shown in the figure.\\

\begin{figure}[h!]
\includegraphics[width=0.8\columnwidth]{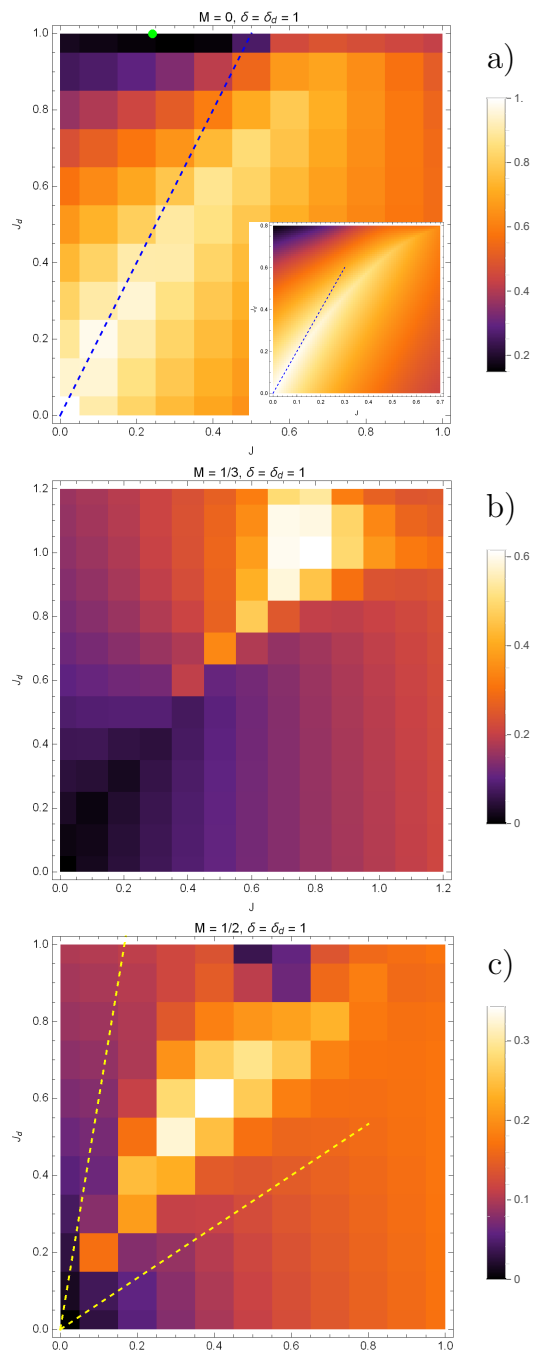}
\caption{(Color online). Phase diagrams of the plateaus at $M = 0$, $1/3$ and $1/2$ (top to bottom) for $ \delta = \delta_d = 1 $ in the plane $ J-J_d $, up values of the order of one, analyzed by means of DMRG.
The color indicates a decreasing plateau width (from white to black). The blue dotted line in the top panel indicates the line of maximum frustration $ g = 0 $.
In the inset of Panel (a) the SE triplet gap is depicted showing good agreement with the DMRG determination.
}
\label{fig:gap-phases}
\end{figure}

On the other hand, Fig \ref{fig:gap-phases} shows the extension of the plateaus at $M = 0$ (a), $1/3$ (b) and $1/2$ (c),  in the plane $ J-J_d $ calculated on finite systems with the DMRG technique. The determination is not intended to be quantitatively accurate, for which an analysis of a finite-size scaling that goes beyond the objectives of the work should be carried out, but rather to determine the regions in the parameters space where the plateaus are most prominent.\\

Let us first consider Fig \ref{fig:gap-phases}(a) where the width of $M=0$ plateau is considered. The largest plateau width ($\Delta =1$) is at the origin, corresponding to the limiting case of isolated dimers (see Fig. \ref{fig:ladder}).
The maximum width of the plateau follows the line of maximum frustration, $ g = 0 $, indicated in this figure with dotted blue line, up to $ J \simeq 0.2 $, beyond which deviations towards $g<0$ become increasingly marked. In the lower right inset of Fig. \ref{fig:gap-phases} (a) the SE triplet gap is depicted showing the same trend as the DMRG determination. This deviation is consistent with the observed shift of the gap peak in Fig. \ref{fig:gap-BO}, predicted by SE and DMRG, as well as the tendency to collapse into a single $g-$dependent triplet gap curve for $g<0$.\\ Another particular case of our model is represented by the point $ (J = 0, J_d = 1) $, corresponding to the homogeneous Heisenberg chain, which is gapless. Note that both DMRG and SE predict a tendency toward the gap closure (dark zone) approaching that point. The model remains gapless along the line $J_d=1$ up to the point $(J \simeq 0.241, J_d =1)$, indicated by a green circle in the Fig. \ref{fig:gap-phases} (a)). From this point the 
system is gapped and for $ (J = 0.5, J_d = 1) $ (Majumdar-Gosh point) the model displays the exact dimer-product zero field ground state, which extends all over the line $ g = 0 $ analyzed in Section \ref{sec:exact-lines}.\\

Regarding the $ M = 1/3 $ plateau, the Fig. \ref{fig:gap-phases} (b), shows that this plateau emerges in a certain reduced and highly frustrated chain limit of the model, where $ J, J_d$, are of the order of the unity, reaching values of plateau width of $ \approx 0.6$. The nature of this plateau is completely different to the others analyzed in this work, having a classical origin \cite{NN-Ising72}, characterized by an "up-up-down" ordering which stabilizes this magnetization value. This state adiabatically evolves from the Ising chain with first and second neighbors interactions, and survives up to the Heisenberg isotropic limit. This plateau has been extensively analyzed in the works of Refs. \onlinecite{Okunishi-Tonegawa03PRB,Okunishi-Tonegawa03JPSJ,AndreasM1/3,Tonegawa04}. In addition, the techniques of strong plaquette expansions that we will use (see next Section) are not specially  suitable to treat this plateau, since it emerges in a region where all the couplings are of the same order. For these 
reasons we will not elaborate more on the same in this work.\\

Finally the panel (c) of Fig. \ref{fig:gap-phases} depicts the extension of the $M=1/2$ plateau determined by DMRG. The emergence of this plateau has been analyzed in terms of a dimerized and frustrated ladder from a numerical point of view \cite{Tonegawa-ladder-98, Tonegawa04}. The dashed yellow lines indicate estimations of high field $(h>1)$ limit of $M=1/2$ phase plateau, from an effective low energy dimer model description of the system \cite{Mila-ladder-98,Sen-ladder-99}, indicating a similar trend to our findings.\\

In the following Sections we will develop a low energy plaquette effective model, which will account for present plateaus (except $M = 1/3$, wich would require higher order terms) and predict the emergence of others at $M = 1/4$ and $3/4$ due to the interplay between frustration and inter-plaquette dimerization.

\subsection{Effective model description of dimerized plaquette Hamiltonian  }
\label{sec:effectiveH}

In the previous Sections we have considered the frustrated homogeneous plaquette case, where the frustration
degree is controlled by  $g = J_d - 2 J $, being maximal along the line $g=0$, where the model exhibits an
exact rung-dimer product state. In this Section we will explore the effect of plaquette dimerization
via the parameters $\delta = K/J$ and $\delta_d=K_d/K$. The interplay between on-plaquette frustration and
inter-plaquette dimerization gives rise to a richer plateaus structure that we will analyze in a weak interacting  plaquettes regime, starting from the limit of decoupled plaquettes. 
It is worth mentioning that an equivalent
model, although with a different parametrization has been studied analytically using a field theory bosonization approach \cite{Cabra-Grynberg-00}. This FLuttstudy, however, is complementary to ours, since it starts from the strong chain limit.\\

We start by analyzing the Hilbert space corresponding to isolated plaquettes, which is spanned by a basis of sixteen states containing two singlets, nine triplets and five quintuplets, which we identify generically by $|s_i\rangle, |t_j\rangle $ and $|q_k\rangle$, respectively. This states are listed in the Appendix \ref{sec:Effective Hamiltonians} (Table \ref{tab:energies}).
On each plaquette in the absence of magnetic field $h$ the ground state corresponds to the singlet $|s_0\rangle$. When the magnetic field is turned on the energy corresponding to the singlet and a triplet becomes closer until
a critical value where a level crossing occurs and a triplet becomes the ground state. By further increasing the magnetic field we obtain a second level crossing between triplet and the  quintet state $|q_0\rangle$. Here the local structure of the plaquettes couplings becomes important. According to the values of the internal couplings two different types of triplets are involved in the crossings, depending on whether $J_d< \frac{2J}{1+J}$ or $J_d>  \frac{2J}{1+J}$. In the following we refer these two different scenarios as cases A and B, respectively.
In the case A (B) the singlet state $|s_0\rangle$ is degenerate with a triplet state $|t_{A(B)}\rangle$ in the first transition, at the critical field $hc_{1A(B)}$, whereas $|t_{A(B)}\rangle$ is degenerate with the quintet state $|q_0\rangle$ in the second transition at the critical field $hc_{2A(B)}$. These possibilities are represented schematically in eq.(\ref{eq:casesAB})
 \begin{eqnarray}
 \label{eq:casesAB}
    |s_0\rangle \xrightarrow{hc_{1A}} |t_A\rangle \xrightarrow{hc_{2A}} |q_0\rangle, \quad J_d < \frac{2J}{1+J}, \nonumber \\
    |s_0\rangle \xrightarrow{hc_{1B}} |t_B\rangle \xrightarrow{hc_{2B}} |q_0\rangle, \quad J_d > \frac{2J}{1+J}.
 \end{eqnarray}
Around these level crossing points we can derive an effective model. Let us write the Hamiltonian as follows
 \ba
 \mathbf{H_\mu}=\mathbf{H}_{0,\mu}+\mathbf{H}_{int,\mu},
 \ea
where the extra index $\mu$ indicates the possible cases mentioned, \emph{ie} $\mu=\{1A,2A,1B,2B\}$, being
 \ba
\mathbf{H}_{0,\mu}&=&\sum_{n} \Big[ \left(\spin_{1,n}\cdot\spin_{2,n}+\spin_{3,n}\cdot\spin_{4,n}\right) \\\nn
&+&J \left(\spin_{1,n}\cdot\spin_{3,n}+ \spin_{2,n}\cdot\spin_{4,n}\right)+J_{d} \, \spin_{2,n}\cdot\spin_{3,n} \\\nn
&-&  hc_\mu\sum_{j=1}^{4}\spin_{j,n} \Big], \\\nn
\mathbf{H}_{int,\mu}&=&\sum_{n}\Big[ K\left( \spin_{3,n}\cdot\spin_{1,n+1}+\spin_{4,n}\cdot\spin_{2,n+1}\right)\\\nn
&+&K_{d}\, \spin_{4,n}\cdot\spin_{1,n+1}-(h-hc_\mu)\sum_{j=1}^{4}\spin_{j,n}\Big].
\ea
We construct the effective Hamiltonian via degenerate perturbation theory \cite{Totsuka1,Mila-ladder-98,Sen-ladder-99,mila-10,4-tube-3},
\ba
\mathbf{H}_{\text{eff},\mu}=\mathbf{H}_{\text{eff},\mu}^{(1)}+\mathbf{H}_{\text{eff},\mu}^{(2)}+\cdots,
\ea
where the superscript indicates the order of the perturbation, although for our purpose in this paper we will only consider the first order of the expansion.

Since at the crossing point the ground state of each plaquette is doubly degenerate (except for $J_d = \frac{2J}{1+J}$ which is triply degenerate and will not be considered here), the original dimerized plaquette model with sixteen states per site is reduced to an \emph{effective} spin-1/2 chain model involving the two mentioned degenerate low energy states at the crossover.

The first order effective Hamiltonian around $hc_\mu$ is obtained by applying the standard degenerate perturbation theory

\begin{equation}
\label{eq:O1-formula}
  \mathbf{H}_{\text{eff},\mu}^{(1)}=\sum_{i,j=1,2}|v_i\rangle\langle v_i| \mathbf{H}_{\text{int},\mu}|v_j\rangle\langle v_j|,
\end{equation}
where states $|v_i\rangle$ and $|v_j\rangle$ span the set $\{|s_0\rangle, |t_{A(B)\rangle}\}$ and
$\{|q_0\rangle, |t_{A(B)\rangle}\}$ for $\mu=1A(B)$ and $\mu=2A(B)$, respectively. In this way is obtained,
up to a constant term,
\ba
\nn
\mathbf{H}_{\text{eff},\mu}^{(1)}&=&\sum_{n}J_{xy,\mu}  \left(\mathbf{S}^{x}_{n}\cdot\mathbf{S}^{x}_{n+1}+ \mathbf{S}^{y}_{n}\cdot\mathbf{S}^{y}_{n+1}\right)  \\
\label{eq:Heff1}
&+&J_{zz,\mu}\mathbf{S}^{z}_{n}\cdot\mathbf{S}^{z}_{n+1}-\tilde{h}_\mu\mathbf{S}^{z}_{n},
\ea
where effective couplings $J_{xy,\mu}, J_{zz,\mu} $ and $\tilde{h}_\mu$ are functions of the original plaquette model couplings $J, J_d, K, K_d$ and $h$. The explicit expressions of these functions are available in the Appendix \ref{sec:Effective Hamiltonians} (Eqs. \ref{eff-J-b}-\ref{eff-J-e}).
In addition, the pseudo spin$-1/2$ operators in Eq. \ref{eq:Heff1} are projectors of the degenerate plaquette basis
\begin{eqnarray}
\label{eq:map1}
  \mathbf{S}^z_n &=& \frac{1}{2} \left(|t_{A(B)}\rangle\langle t_{A(B)}|-|s_0\rangle\langle s_0|\right)_n \nonumber \\
  \mathbf{S}^\dag_n &=& \left(|s_0\rangle\langle t_{A(B)}|\right)_n, \quad \mu = 1A(B),
  \end{eqnarray}
\begin{eqnarray}
\label{eq:map2}
  \mathbf{S}^z_n &=& \frac{1}{2} \left(|q_0\rangle\langle q_0|-|t_{A(B)}\rangle\langle t_{A(B)}|\right)_n \nonumber \\
  \mathbf{S}^\dag_n &=& \left(|t_{A(B)}\rangle\langle q_0|\right)_n, \quad \mu = 2A(B),
\end{eqnarray}
where $\mathbf{S}^\dag_n=\mathbf{S}^x_n + i \mathbf{S}^y_n $.

\subsection{Phase diagram of the spin$-1/2$ chain effective Hamiltonian}
\label{sec:effectiveH}

\begin{figure}[t]
\includegraphics[width=1.0\columnwidth]{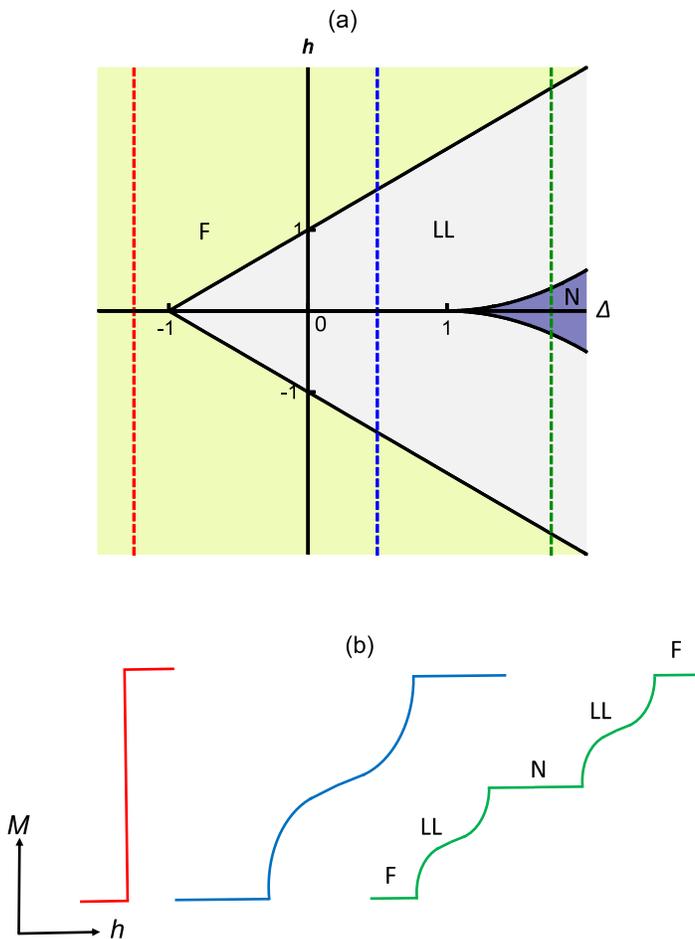}
\caption{(Color online). Panel (a): Phase diagram $h$ vs $\Delta$ (in units of $J_{x,y}$) of the XXZ spin$-\frac12$ chain. The three phases present in the model are ferromagnetic (F, yellow), N\'eel (N, blue) and Luttinger liquid (LL, gray). The F and N phases are gapped, and exhibit spin$-1$ magnon and spin$-1/2$ spinon excitations, respectively, whereas LL phase is gapless and presents quasi-long range order. Panel (b) Schematic representation of the magnetization curves $M(h)$ corresponding to the three paths indicated with red, blue and green dotted lines in panel (a). Note that each gapped phase translates into plateaus in the magnetization curves(see the text for details).}
\label{fig:BA-phases}
\end{figure}

The effective model given by Eq.(\ref{eq:Heff1}) corresponds to an XXZ spin$-\frac12$ chain and can be solved exactly via the Bethe ansatz \cite{bethe}. Here we briefly review the main characteristics of the Bethe ansatz solution, considering the different phases present, and their implications in the magnetization process.
For simplicity in this discussion we will simplify the notation of Eqs.(\ref{eq:Heff1})by removing supra- and sub- indices and indicating the three principal parameters of the model as $J_{xy}$, $ J_{zz} $ and the magnetic field $h$. The phase diagram in the plane $( \Delta =J_{zz}/J_{xy}, h/J_{xy})$ is shown
in the Fig.(\ref{fig:BA-phases},a). As can be observed there are three phases: Ferromagnetic (F, yellow), N\'eel (N, blue) and Luttinger Liquid (LL, gray). The F and N phases are gapped, with spin$-1$ \emph{magnon} and spin$-1/2$ \emph{spinon} domain wall type of excitations, respectively. On the other hand, the intermediate phase LL is gapless and exhibits quasi-long range order.
Regarding critical lines in Fig. \ref{fig:BA-phases}(a), the straight lines separating LL and F phases are explicitly given in terms of $\Delta$  by
\begin{equation}
\label{eq:BA-L_F}
 h_{L-F}=\pm(1+\Delta),
\end{equation}
whereas the border between LL and N phase is obtained by solving \cite{chain-in-field}
\begin{equation}
\label{eq:BA-L_N}
  h_{L-N}=\pm\sinh (g)\sum_{n=-\infty}^{\infty}\frac{(-1)^n}{\cosh(n g)},
\end{equation}
where $g=\operatorname{arcosh}\Delta$. Note that both, $h_{L-F}$ and $h_{L-N}$, are given in units of $J_{xy}$.

The different phases present in the model translate into distinctive characteristics in the magnetization curves. The gapped phases F and N exhibit a plateau in the magnetization curve, while in the LL phase the magnetization continuously increases with the magnetic field. This dependence is illustrated in Fig.\ref{fig:BA-phases}. In panel (a) of this figure three paths with red, blue and green dotted lines are indicated in the phase diagram, while in Fig. \ref{fig:BA-phases}(b) the magnetization curves corresponding to these three paths are showed schematically. Note that the vertical axis in panel (a) is represented as horizontal axis in panel (b). The red curve represents the simplest case, where the system is always in the F phase and has a stepwise structure with two plateaus, jumping directly from one to another with the sign change of the magnetic field. The blue magnetization curve represents an intermediate case, in which the two plateaus corresponding to the F phase are connected through a region 
of continuous growth of the magnetization with the field, whose shape is characteristic of the LL phase. Finally, the green magnetization curve in Fig. \ref{fig:BA-phases}(b) indicates the more general case, which comprises the three phases. In this case, in addition to the plateaus at the extremes due to the F phase, an additional intermediate plateau indicative of the N phase is present. These three plateaus are connected by LL phases, consistently with the path indicated in green in Fig. \ref{fig:BA-phases}(a).\\
Before concluding with this Section we would like to comment on another case where the plateau structure has a simple stepped form. This case is the \emph{Ising} limit of the model, in which $J_{x,y}\rightarrow 0\, (\Delta \rightarrow \infty)$, \emph{ie} the right infinite end of Fig. \ref{fig:BA-phases}(a). Here the magnetization curve exhibits the three plateaus with direct jumps between them, without intermediate LL phase.

\subsection{Effective model description of magnetization process}
In this Section we will analyze how the effective low-energy spin-1/2 chain model with anisotropy $\Delta$, obtained previously, captures the main aspects of the magnetization process of the original plaquette model in a certain, strong plaquette coupling, regime of the parameters space. The effective model is not only useful to account for the numerical results obtained through DMRG, but also brings a direct physical interpretation of the structure of plateaus that can emerge in the original model. In addition, it provides a tool that allows us to detect some characteristics, such as small intermediate plateaus, which would be difficult to detect by direct numerical scanning.

To apply the results of the effective model to the original plaquette model, note that according to Eq.(\ref{eq:map1}), the effective pseudo-spin operators eigenvalues $S_z = -1/2 (1/2)$ correspond to plaquette $S_z = 0 (1)$, of the singlet $|s_0\rangle$ and triplet $|t_{A(B)}\rangle$ eigenvalues of the first level crossing $1A(B)$. Therefore the three possible magnetization plateaus in the effective chain model at $\{-1,0,1\}$ are mapped onto $\{0,1/4,1/2\}$ magnetization plateaus in the plaquette model, normalized to maximum spin per site and plaquette, respectively. Similarly for the second level crossing $2A(B)$ between the triplet $|t_{A(B)}\rangle$ and quintet $|q_0\rangle$, the Eq.(\ref{eq:map2}) maps $\{-1,0,1\}$ onto $\{1/2, 3/4,1\}$, corresponding to effective chain and plaquette models, respectively.

Now we will describe qualitatively the structure of the magnetization curves in terms of the effective models. To this end, the Fig. \ref{fig:mag-disp} illustrates different representative plateaus structures in the original model obtained by means of DMRG, showed with blue solid line. In particular, the Fig. \ref{fig:mag-disp}(a) shows magnetization curves with a single transition between type-F phases at zero magnetic field and saturation, intermediated by a LL phase. Note that the structure of small steps in the LL portion of the curve is due to the finite size of the numerical calculation. In Fig.\ref{fig:mag-disp}(b) an additional plateau emerges at mid-magnetization. However, according to the previous analysis, this plateau is type-F, like the other two present here. The cases of Figs. \ref{fig:mag-disp}(a,b) are not dimerized, \emph{ie}  $\delta=\delta_d=1$, and illustrate a general feature: in absence of dimerization between plaquettes there can only be plateaus at $ M = 0,1 / 2 $ and $ 1 $ in the 
magnetization curve, sharing spin$-1$ magnon excitations.  The origin of these plateaus is intrinsic to each isolated plaquette, as a consequence of the two (1A(B), 2A(B)) possible level crossings, which are only renormalized by the interaction between plaquettes.
\\

On the other hand, the Fig. \ref{fig:mag-disp}(c) show a plaquette-dimerized case, where additional plateaus at $M = 1/4, 3/4$, are present. These plateaus are indicative of the N-type phase which, as we have mentioned,  present spinon-like excitations and are generated by the interaction between plaquettes. A necessary (but not sufficient) condition for their emergence is the presence of plaquette dimerization.\\
Regarding the size of type-N plateaus, we have observed that in general are smaller than those of type-F.  For this reason the effective model is very useful for their detection, since numerically they could easily be overlooked, given the quite large number of parameters involved in the original model.\\
In addition, the effective model not only facilitates the detection of the intermediate N-type plateaus, but also highlights the role of frustration in the emergence of them. In fact it is observed that apart from dimerization, type-N plateaus grow with frustration,
being largest along the line of maximum frustration, \emph{ie} $ g=J_d-2J=0 $ and $ g_k=K_d-2K=0 $, whenever this condition can be satisfied, which depends on the crossing levels (A or B) involved. It is interesting to note that in a certain range of validity frustration-induced type-N plateaus are maximal along the maximum frustration line, where in addition the original model (Eq.(1)) exhibits an exact dimer singlet-product zero field ground state. In some sense it suggests that this condition is not only relevant at zero magnetic field, but also play a role in the magnetization process.

Finally, regarding the dotted red and yellow vertical lines showed in the three panels of Fig. \ref{fig:mag-disp}, they are quantitative estimations of critical fields delimiting the different plateaus based on the dispersion of magnon and spinon excitations of the effective model, and will be discussed in the next Section.



\subsection{Magnetic excitations and critical fields}

Here we will analyze the excitations on the plateau structures in the effective chain model, which in turn will allow us to obtain the critical fields, \emph{ie} the borders of the plateaus and compare with the numerical DMRG results 
from the original model. \\First we consider the saturation plateaus at $-(+1)$ in the effective chain model. These correspond to ferromagnetic states, with all the effective spins on each site polarized along the negative and positive direction of the effective field, respectively. The elemental magnon excitation consists of a spin-wave composed by a linear combination of states with a local one-spin inversion, carrying $ \Delta S_z =+(-1) $. The application of the effective Hamiltonian (Eq.(\ref{eq:Heff1})) on this state (subtracting the ground state) gives rise, by a standard calculation, to the magnon (M) dispersion
\begin{equation}
\label{eq:magnon-disp}
  \omega^\pm_{M,\mu}(k)=J_{xy,\mu}\cos(k)-J_{zz,\mu}\mp \tilde{h}_\mu.
\end{equation}
Eq.(\ref{eq:magnon-disp}) describes all magnon-type of excitations present in the strong-plaquette expansion of the original model, via the mapping given by Eqs.(\ref{eq:map1},\ref{eq:map2}), and the effective couplings ((Eqs. \ref{eff-J-b}-\ref{eff-J-e})), namely (\emph{i}): right (high field) borders of $M = 0 \quad (+, \mu = 1A (B))$ and $M=1/2 \quad (+, \mu = 2A (B))$; (\emph{ii}): left (low field) boundaries of $M = 1/2 \quad (-, \mu = 1A (B))$ and $M = 1 \quad (-, \mu = 2A (B))$.

Let us now consider the $M=0$ plateau in the effective model, which is associated to a double degenerate N\'eel state. Here the elemental spinon excitation consists of a spin-wave composed by a linear combination of domain wall states with two consecutive spin inversions, carrying $ \Delta S_z =+(-1/2) $. Following a similar procedure as the magnon case we obtain the spinon (S) dispersion
\begin{equation}
\label{eq:spinon-disp}
  \omega^\pm_{S,\mu}(k)=J_{xy,\mu}\cos(2k)+\frac{1}{2}J_{zz,\mu}\mp \frac{1}{2} \tilde{h}_\mu.
\end{equation}
Equivalently to the previous case, Eq.(\ref{eq:spinon-disp}) describes all spinon-type of excitations, namely (\emph{i}): right (high field) borders of $M = 1/4 \quad (+, \mu = 1A (B))$ and $M=3/4 \quad (+, \mu = 2A (B))$; (\emph{ii}): left (low field) boundaries of $M = 1/4 \quad (-, \mu = 1A (B))$ and $M = 3/4 \quad (-, \mu = 2A (B))$.

\begin{figure}[t]
\includegraphics[width=0.8\columnwidth]{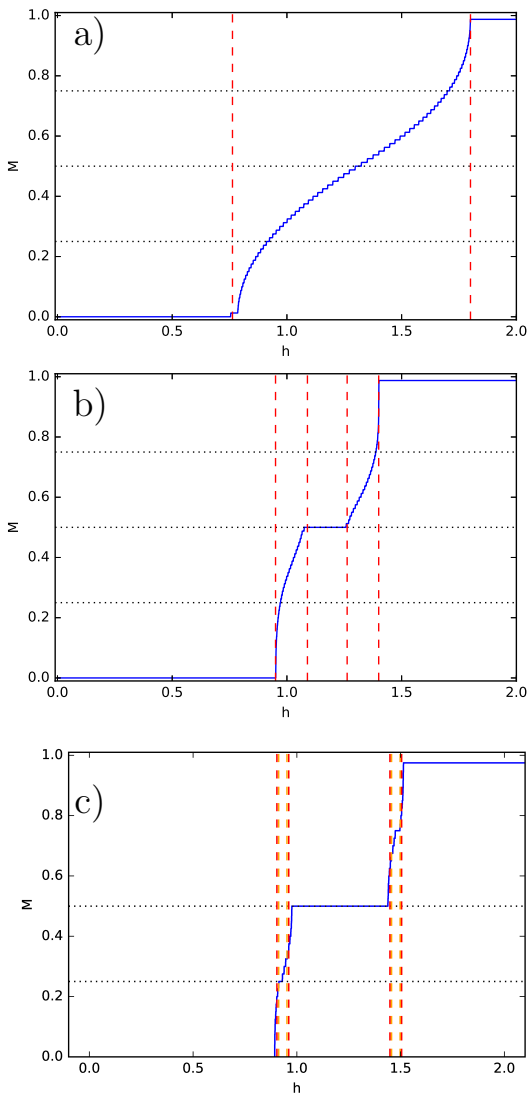}
\caption{(Color online). Magnetization curves corresponding to $J=K=0.4$ and $J_{d}=K_{d}=0.3$ (panel a)), $J=K=0.2$ and $J_{d}=K_{d}=0.3$ (panel b)), $J=0.8$, $J_{d}=0.6$, $K=0.2$  and $K_{d}=0.45$ (panel c)).
Red-dashed lines and orange-dashed lines correspond to critical fields obtained by means Magnon and Spinon dispersions respectively. 
}
\label{fig:mag-disp}
\end{figure}

From the above magnon $\omega^\pm_{M,\mu}(k)$ and spinon $\omega^\pm_{S,\mu}(k)$ dispersions we can calculate the critical fields, \emph{ie} the edges and the width of the plateaus present in the model. For this, we impose the condition of gap closure (minimum of the dispersion), which allows to determine the effective fields in terms of the effective couplings and thus the critical fields according to the original couplings of the model. Due to the number of parameters of the model, in general we obtain the critical fields numerically which is much simpler compared to the  DMRG or the numerical solution of the Bethe-ansatz equations approaches. \\
The techniques used in this work to determine the critical fields: DMRG, and effective models involve different types of approximation. DMRG suffers finite size effects and the effective models have a perturbative origin. 
However, these methods are somewhat complementary and provide consistent results. Although, as we have mentioned, it is not the aim of this work to quantitatively determine the edges of each plateau in the whole parameter
space of the model, it is interesting to compare the results of the different techniques in some representative cases. For this purpose, in Fig. \ref{fig:mag-disp} we show the critical fields determined 
by DMRG (full magnetization curve in blue line) and the edges of the different plateaus present by applying the gap closure condition of the corresponding magnon dispersion (red-dashed lines) or spinon dispersions (orange-dashed lines).
As can be observed there is a very good agreement between the different techniques, which is maintained up to intermediate couplings (compared with $J_0$).\\

As previously mentioned and as can be seen in the Fig. \ref{fig:mag-disp}(c) the plateaus at $M = 1/4$ and $3/4$ are small compared to the others, so that they could be difficult to detect numerically by sweeping a fairly large parameter space as in this model using DMRG. It is therefore of particular interest to be able to estimate not only the presence but the maximum possible size of these plateaus as well as their interdependence with frustration and dimerization.\\
From this point of view the effective model offers a direct way of evaluating positions and maximum widths of the plateaus. This will occur when there is no intermediate Luttinger liquid phase between plateaus, \emph{ie} the limit of the effective Ising model where $J_{xy,\mu} = 0$. In this condition the magnon and spinon disperson bands do not propagate (they are flat) and the edges of fields predicted by both merge. From the Eqs (\ref{eq:magnon-disp},\ref{eq:spinon-disp}) we see that the effective critical fields at the Ising limit satisfy
\begin{equation}\label{hc-Ising}
  J_{zz,\mu}=\pm \tilde{h}_\mu, \quad \textmd{for} \quad  J_{xy,\mu}=0.
\end{equation}
Note in passing that this condition is obtained from the effective model exact line $h_{L-F}$  (Eq.(\ref{eq:BA-L_F}))
in the limit $\Delta \rightarrow \infty$.\\

By solving the Eqs.(\ref{hc-Ising}) in terms of the original variables of the model (Eqs. \ref{eff-J-b}-\ref{eff-J-e})) we get the following cases for the widest plateaus $\Delta h$. For case A $(J_d<\frac{2J}{1+J})$, the Ising limit is obtained for $g_k=K_d-2K=0$ with the same $\Delta h=K/2$ for both $M=1/4$ and $M=3/4$ plateaus. \\
On the other hand, for case B $(J_d>\frac{2J}{1+J})$, the plateau widths are not the same, although we can also obtain analytical expressions for both. In the case $M = 1/4$ the Ising condition is obtained for both $g=J_d-2J=0$ and $g_k=K_d-2K=0$ although these conditions are not satisfied for $M = 3/4$. \\


\section{Conclusions}
\label{sec:conclusions}

In the present paper we have studied the magnetization process in a zig-zag quantum antiferromagnet in the presence of both frustration and plaquette dimerization. Due to the complexity of the unit cell, several cases are present giving a rich structure of the magnetization curve. At zero magnetic field the ground state is exactly determined under the conditions $g=J_d-2J=0$ and $g_k=K_d-2K=0$ and corresponds to a product state of spin singlets.
Around this highly frustrated line we investigate the $M=0$ plateau by using dimer series expansions, bond operators mean field theory and numerically by means of density matrix renormalization group.

In order to analyze the magnetization process we complement the numerical results with first order low energy effective models starting from the limit of decoupled plaquettes. From a qualitative point of view, the effective models capture the essential features obtained numerically and bring a simple physical interpretation of the emergent structure of plateaus in the original zig-zag model. In addition, the analysis of magnon and spinon excitations in the efective models allows us to obtain estimations of the critical fields bordering the plateaus, which are in a very good quantitative agrement with the numerical computations
in a strong plaquette dimerization regime. 

Our study also suggest that the combined effect of on-plaquette frustration, controlled by $g$ and $g_k$ and inter-plaquette dimerization, via $\delta=K/J$ and $\delta_d=K_d/J_d$, plays a central role in the structure of the magnetization curve. In fact whenever is possible, according to the values of the couplings involved, $M = 1/4$ and $3/4$ plateaus are widest along the line of maximum frustration. In this sense we say that these plateaus are induced by frustration, which together with the dimerization between plaquettes, provide a rich interplay between both aspects that are intrinsic to the model. For this reason, the maximally frustrated $g$ and $g_k = 0$ line is not only fundamental to zero field, but plays a central role beyond zero magnetic field, which is reflected in its influence on the resulting magnetization process.

\section*{Acknowledgments}
We acknowledge useful discussions with M. Matera.
C. A. Lamas is supported by CONICET (PIP 1691) and ANPCyT (PICT 2013-0009)

\appendix

\section{Effective Hamiltonians}
\label{sec:Effective Hamiltonians}
As we discuss in the main body of the text, we construct an effective Hamiltonian via degenerate perturbation theory.
The energy corresponding to the states on a square plaquette are presented in Table \ref{tab:energies}.
By applying a standard degenerate perturbation theory around the points $hc_\mu$ where the lowes eigenvalues become degenerate we obtain, up to a constant term, a first order effective Hamiltonian in the form
\ba
\nn
\mathbf{H}_{\text{eff},\mu}^{(1)}&=&\sum_{n}J_{xy,\mu}  \left(\mathbf{S}^{x}_{n}\cdot\mathbf{S}^{x}_{n+1}+ \mathbf{S}^{y}_{n}\cdot\mathbf{S}^{y}_{n+1}\right)  \\
\label{eq:Heff1-App}
&+&J_{zz,\mu}\mathbf{S}^{z}_{n}\cdot\mathbf{S}^{z}_{n+1}-\tilde{h}_\mu\mathbf{S}^{z}_{n},
\ea
where the couplings are given by
\begin{widetext}

\small
\ba
\label{eff-J-b}
J_{xy,1A}&=&-\frac{\left(\sqrt{4 J^2-2 J (J_d+2)+J_d^2-2
   J_d+4}+J-J_d+1\right) (2 K-K_{d})}{6
   \sqrt{4 J^2-2 J (J_d+2)+J_d^2-2 J_d+4}}\\
J_{zz,1A}&=&\frac{1}{16} (2 K+K_d)\\
\tilde{h}_{1A}&=&h+\frac{1}{2} \left(-\sqrt{4 J^2-2 J J_{d}-4 J+J_{d}^2-2
   J_{d}+4}-J_{d}\right)+\frac{1}{16} (2
   K+K_{d})\\
   %
   %
J_{xy,1B}&=&\frac{K_{d} \left(\gamma  (1-2 \rho )-3 J^2+J (\gamma +2 J_{d}+6)-2 J_{d}^2-J_{d} (\gamma +\rho -2)\right)+2 (J-1) K (-3 \gamma +J-J_{d}-2 \rho +1)-3
   K_{d}}{12 \gamma  \rho }\\
J_{zz,1B}&=&\frac{1}{16} \left(\frac{J_{d}^2 (K_{d}-2 K)}{(J-1)^2+J_{d}^2}+\frac{2 J_{d} K_{d}}{\sqrt{(J-1)^2+J_{d}^2}}+2 K+K_{d}\right)\\
\tilde{h}_{1B}&=&h+\frac{1}{2} \left(\sqrt{J^2-2 J+J_{d}^2+1}-\sqrt{4 J^2-2 J J_{d}-4 J+J_{d}^2-2 J_{d}+4}-J-1\right)\\
&+&\frac{1}{16} \left(-\frac{J_{d}^2 (K_{d}-2
   K)}{(J-1)^2+J_{d}^2}-\frac{2 J_{d} K_{d}}{\sqrt{(J-1)^2+J_{d}^2}}-2 K-K_{d}\right)\\
   %
   %
J_{xy,2A}&=&\frac{K_{d}}{4}-\frac{K}{2}\\
J_{zz,2A}&=&\frac{K}{8}+\frac{K_{d}}{16}\\
\tilde{h}_{2A}&=&h-J-2 \left(-\frac{3 K}{16}-\frac{3 K_{d}}{32}\right)-1\\
   %
   %
J_{xy,2B}&=&\frac{K_{d} \left(J_{d}-\sqrt{(J-1)^2+J_{d}^2}\right)-2 (J-1) K}{4 \sqrt{(J-1)^2+J_{d}^2}}\\
J_{zz,2B}&=&\frac{1}{16} \left(\frac{J_{d}^2 (K_{d}-2 K)}{(J-1)^2+J_{d}^2}-\frac{2 J_{d} K_{d}}{\sqrt{(J-1)^2+J_{d}^2}}+2 K+K_{d}\right)\\
\tilde{h}_{2B}&=&h+\frac{1}{2} \left(-\sqrt{J^2-2 J+J_{d}^2+1}-J-J_{d}-1\right)+\frac{1}{16} \left(\frac{J_{d}^2 (K_{d}-2 K)}{(J-1)^2+J_{d}^2}+\frac{2 J_{d}
   K_{d}}{\sqrt{(J-1)^2+J_{d}^2}}-3 (2 K+K_{d})\right)\\
\ea
\normalsize
where
\ba
\gamma&=&\sqrt{(J-1)^2+J_{d}^2}\\
\label{eff-J-e}
\rho&=&\sqrt{4 J^2-2 J (J_{d}+2)+(J_{d}-2) J_{d}+4}\\
\ea

\end{widetext}

\section{Dimer series expansion}
In this Section we display the explicit expressions for ground state energy and triplet dispersion
employed in the work, for the case of homogeneous plaquettes \emph{ie} $\delta=\delta_d=1$. The method employed
is the continuous unitary transformation (CUT) \cite{SE-CUT}. Here for brevity we present explicit results to $O (6)$ although the calculations shown have been made using $O (10)$.
The ground state energy (per dimer) reads
\begin{widetext}
\small
\ba
\nn
 \textrm{egs} &=& -\frac{3}{4}-\frac{3 g^2}{32}-\frac{3 g^2 J}{32}
  -\frac{3 g^3}{128}-\frac{g^2 J^2}{8}-\frac{5 g^3 J}{64} -\frac{13 g^4}{2048}
  -\frac{17 g^2 J^3}{96}-\frac{73 g^3J^2}{384}-\frac{89 g^5}{24576}-\frac{289 g^4 J}{6144} -\frac{155 g^2 J^4}{576}\\
&-&\frac{47 g^4 J^2}{256} -\frac{2297 g^5 J}{73728}-\frac{463 g^6}{196608} ,
\ea
\normalsize

\end{widetext}
where $g=J_d-2 J$. Note that for $g=0$ only the first term $-\frac{3}{4}$ survives, corresponding to the
dimer-product exact ground state analyzed in Section I. Similarly the triplet dispersion is given by $\omega_{SE}=\sum_{n=0}^\infty c_n \cos(n k),$
where $c_n$ coefficients up to O(6) are
\begin{widetext}
 \small
\ba
\nonumber
c_0 &=& \textcolor[rgb]{0.00,0.00,1.00}{1}-\frac{g^2}{16}+\frac{3 g^3}{64}+\frac{23 g^4}{1024}-\frac{3g^5}{256}+\frac{1273 g^6}{221184}
-g J-\frac{g^2 J}{16}
+\frac{23 g^3 J}{64}-\frac{43 g^4 J}{512}-\frac{71 g^5 J}{1536}-J^2-g J^2
+\frac{55 g^2 J^2}{64} \\ \nonumber
&+&\frac{57 g^3J^2}{256}-\frac{55121 g^4 J^2}{55296}-J^3+\frac{gJ^3}{2}+\frac{105 g^2 J^3}{64}
-\frac{9239 g^3J^3}{2304}-\frac{J^4}{8}+\frac{99 g J^4}{32}-\frac{37399g^2 J^4}{6912}+\frac{7 J^5}{4}
-\frac{361 g J^5}{576}+\frac{367 J^6}{192},\\
\nonumber
c_1 &=& \textcolor[rgb]{0.00,0.00,1.00}{-\frac{g}{2}}-\frac{g^2}{4}+\frac{g^3}{32}+\frac{5g^4}{256}-\frac{35 g^5}{2048}
+\frac{3121 g^6}{221184}-g J-\frac{g^2 J}{4}
+\frac{19 g^3 J}{64}-\frac{337 g^4 J}{1536}-\frac{227 g^5 J}{13824}-J^2-g J^2 +\frac{15 g^2 J^2}{16} \\ \nonumber
&-&\frac{45 g^3 J^2}{128}-\frac{92525 g^4 J^2}{55296}-J^3+\frac{9 g J^3}{8}+\frac{25 g^2
J^3}{16}
-\frac{2863 g^3 J^3}{384}+\frac{J^4}{4}+\frac{881 g J^4}{192}-\frac{11815 g^2 J^4}{1152}
+\frac{23 J^5}{8}
-\frac{1165 g J^5}{576}+\frac{133 J^6}{48},\\
\nonumber
c_2 &=& \textcolor[rgb]{0.00,0.00,1.00}{-\frac{g^2}{16}}-\frac{g^3}{32}-\frac{15 g^4}{512}-\frac{283 g^5}{18432}+\frac{79 g^6}{16384}
-\frac{g^2 J}{8}-\frac{5 g^3 J}{32}
-\frac{1405 g^4 J}{4608}-\frac{1679 g^5 J}{36864}+
\frac{g^2 J^2}{64}-\frac{1243 g^3 J^2}{1152} -\frac{47249 g^4 J^2}{36864}\\ \nonumber
&+&\frac{5 g J^3}{8}-\frac{19 g^2 J^3}{32}-\frac{11933 g^3 J^3}{2304}
+\frac{3 J^4}{8}+\frac{67 g J^4}{48}
-\frac{5419 g^2 J^4}{768}+\frac{9 J^5}{8}-\frac{247 g J^5}{96}
+\frac{107 J^6}{192},\\
\nonumber
c_3 &=& \textcolor[rgb]{0.00,0.00,1.00}{-\frac{g^3}{64}}-\frac{g^4}{48}-\frac{9 g^5}{1024}-\frac{337 g^6}{147456}
-\frac{g^3 J}{12}-\frac{203 g^4 J}{1536}
-\frac{1355 g^5 J}{36864}-\frac{5 g^2 J^2}{96}
-\frac{71 g^3 J^2}{192}-\frac{2099 g^4 J^2}{6144}-\frac{17 g^2 J^3}{48}\\ \nonumber
&-&\frac{5137 g^3 J^3}{4608}-\frac{19 g J^4}{192}-\frac{1973 g^2 J^4}{1152}-\frac{113 g J^5}{96}
-\frac{29 J^6}{96},\\ \nonumber
c_4 &=& \textcolor[rgb]{0.00,0.00,1.00}{-\frac{5 g^4}{1024}}-\frac{67 g^5}{9216}-\frac{13373 g^6}{1769472}-\frac{67 g^4 J}{2304}
-\frac{1405 g^5 J}{24576}
-\frac{5 g^3 J^2}{288}-\frac{3643 g^4 J^2}{36864}-\frac{431 g^3 J^3}{13824}
+\frac{11 g^2 J^4}{768},\\ \nonumber
c_5 &=& \textcolor[rgb]{0.00,0.00,1.00}{-\frac{7 g^5}{4096}}-\frac{767 g^6}{221184}-\frac{767 g^5 J}{55296}-\frac{497 g^4 J^2}{55296},\\
c_6 &=& \textcolor[rgb]{0.00,0.00,1.00}{-\frac{21 g^6}{32768}},
\label{eq:coef-SE}
\ea
\normalsize
\end{widetext}

It is interesting to compare previous SE coefficients with the BO-HP expansion
$\omega_{HP}=\sqrt{1-g \cos k}= \sum_{n=0}^\infty \tilde{c_n} \cos(n k) $,
which to O(6) reads
\small
\ba
\nonumber
\tilde {c_0} &=& \textcolor[rgb]{0.00,0.00,1.00}{1}-\frac{g^2}{16}-\frac{15 g^4}{1024}-\frac{105 g^6}{16384},\\ \nonumber
\tilde {c_1} &=& \textcolor[rgb]{0.00,0.00,1.00}{-\frac{g}{2}}-\frac{3 g^3}{64} -\frac{35 g^5}{2048}, \\ \nonumber
\tilde{c_2} &=& \textcolor[rgb]{0.00,0.00,1.00}{-\frac{g^2}{16}}-\frac{5 g^4}{256}-\frac{315 g^6}{32768},\\ \nonumber
\tilde{c_3} &=& \textcolor[rgb]{0.00,0.00,1.00}{-\frac{g^3}{64}}-\frac{35 g^5}{4096}, \\ \nonumber
\tilde{c_4} &=& \textcolor[rgb]{0.00,0.00,1.00}{-\frac{5 g^4}{1024}}-\frac{63 g^6}{16384}, \\ \nonumber
\tilde{c_5} &=& \textcolor[rgb]{0.00,0.00,1.00}{-\frac{7 g^5}{4096}}, \\
\tilde{c_6} &=& \textcolor[rgb]{0.00,0.00,1.00}{-\frac{21 g^6}{32768}}.
\label{eq:coef-BO-HP}
\ea
\normalsize
From Eqs(\ref{eq:coef-SE}) and Eqs(\ref{eq:coef-BO-HP}), we note that bond operators technique is only exact at leading order (blue highlighted terms).
This connection between both methods illustrates the effect of neglecting interactions beyond the quadratic order in the bosonic model (ref Mikeska).

\begin{widetext}
 
\begin{table}[h]
\centering
 \begin{tabular}{|c|c|c|c|}
 \hline
 \small{$S_T$}&\small{$S^z_T$} & & Energy \\
 \hline
 &&&\\
 $0$ & $0$ & $ |s_0\rangle $ & { \blue $- \frac{J}{2} - \frac{J_d}{4} - \frac{1}{2} \sqrt{4 J^{2} - 2 J J_d - 4 J + J_d^{2} - 2 J_d + 4} - \frac{1}{2}$ }\\
 & $0$ & & $- \frac{J}{2} - \frac{J_d}{4} + \frac{1}{2} \sqrt{4 J^{2} - 2 J J_d - 4 J + J_d^{2} - 2 J_d + 4} - \frac{1}{2}$ \\
 &&&\\
 \hline
 &&&\\
  & $-1$ &    $|t_{A}\rangle $    &{\blue $- h - \frac{J}{2} + \frac{J_d}{4} - \frac{1}{2}$ } \\
  & $-1$ &  $|t_{B}\rangle $ & {\blue $- h - \frac{J_d}{4} - \frac{1}{2} \sqrt{J^{2} - 2 J + J_d^{2} + 1}$ }\\
  & $-1$ & & $- h - \frac{J_d}{4} + \frac{1}{2} \sqrt{J^{2} - 2 J + J_d^{2} + 1}$ \\
  & $0$ & & $- \frac{J_d}{4} - \frac{1}{2} \sqrt{J^{2} - 2 J + J_d^{2} + 1}$ \\
 $1$ & $0$ & & $- \frac{J_d}{4} + \frac{1}{2} \sqrt{J^{2} - 2 J + J_d^{2} + 1}$ \\
  & $0$ & & $- \frac{J}{2} + \frac{J_d}{4} - \frac{1}{2}$ \\
  & $1$ & & $h - \frac{J}{2} + \frac{J_d}{4} - \frac{1}{2}$ \\
  & $1$ & & $h - \frac{J_d}{4} - \frac{1}{2} \sqrt{J^{2} - 2 J + J_d^{2} + 1}$ \\
  & $1$ & & $h - \frac{J_d}{4} + \frac{1}{2} \sqrt{J^{2} - 2 J + J_d^{2} + 1}$ \\
  &&&\\
  \hline
  &&&\\
  & $-2$ & $|q_0\rangle $  &{\blue  $- 2 h + \frac{J}{2} + \frac{J_d}{4} + \frac{1}{2}$ }\\
  &  $-1$ && $ - h + \frac{J}{2} + \frac{J_d}{4} + \frac{1}{2}$\\
 $2$ &  $0$ & & $ \frac{J}{2} + \frac{J_d}{4} + \frac{1}{2}$\\
  &  $1$  && $  h + \frac{J}{2} + \frac{J_d}{4} + \frac{1}{2}$\\
  & $2$ && $ 2 h + \frac{J}{2} + \frac{J_d}{4} + \frac{1}{2}$\\
  &&&\\
  \hline
 \end{tabular}
 \caption{\label{tab:energies} Eigenvalues corresponding to the 16 eigenstates of the plaquette. The four low energy states involved in the perturbative calculation are
 labeled as $|s_0\rangle$, $|t_{A}\rangle $, $|t_{B}\rangle $ and $|q_0\rangle $ .}
\end{table}

\end{widetext}

%
\bibliography{refs-J-Jd}

\end{document}